\documentclass[useAMS,usenatbib]{mn2e}
\usepackage{epsfig}

\newdimen\figurewidth
\def\mr{\mathrm}
\def\ledd{{\cal L}_{\mr{Edd}}}
\def\fedd{{\cal F}_{\mr{Edd}}}
\def\feff{{\cal F}_{\mr{eff}}}
\def\func{{\cal W}}
\def\eff{f}

\def\gtrsim{\mathrel{\hbox{\rlap{\hbox{\lower3pt\hbox{$\sim$}}}\raise2pt\hbox{$>$}}}}
\def\lesssim{\mathrel{\hbox{\rlap{\hbox{\lower3pt\hbox{$\sim$}}}\raise2pt\hbox{$<$}}}}

\title[Super Eddington Accretion Disks]{Super Eddington Slim Accretion Disks with Winds}
\author[Dotan \& Shaviv]{Calanit Dotan$^{1}$ and Nir J. Shaviv$^{1}$\\
\noindent
$^{1}$Racah Institute of Physics, Hebrew University of Jerusalem, Jerusalem 91904, Israel }
\begin{document}


\pagerange{\pageref{firstpage}--\pageref{lastpage}} \pubyear{2010}

\maketitle

\label{firstpage}

\begin{abstract}

We construct Super-Eddington Slim Disks models around both stellar and super-massive black holes by allowing the formation of a porous layer with a reduced effective opacity. We show that at high accretion rates, the inner part of the disks become radiation pressure dominated. However, unlike the standard scenario in which the disks become thick, super-Eddington disks remain slim. In addition, they accelerate a significant wind with a ``thick disk" geometry. We show that above about 1.5 times the standard critical mass accretion rate (needed to release the Eddington luminosity), the net luminosity released is above Eddington. At above about 5 times the standard critical rate, the central BH accretes more than the Eddington accretion rate. Above about $20 \dot{m}_{crit}$, the disk remains slim but the wind becomes spherical, and the present model breaks down.   
\end{abstract}


\section{Introduction}

The accretion of matter onto compact objects can often be described using the run-of-the-mill thin accretion disk model of \citet[][S\&S]{Shakura1973}. Because the accretion disk is optically thick, matter can radiate the potential energy it dissipates and remain cold, thus forming a geometriclaly ``thin" disk. The turbulent viscosity responsible for the dissipation is often described through the standard $\alpha$-model, and it is also responsible for the transport of angular momentum outwards. 

The S\&S thin disk model applies to a wide range of conditions found in nature, however, when one of the underlying assumptions break down, so does the model. 

At sufficiently low accretion rates, the disk becomes optically thin and it cannot radiate the energy dissipated. This energy is therefore advected with the flow, forming the so called Advection Dominated Accretion Flow \citep{Ichi:1977,NarayanYi,Abra:95}. Because of the high temperatures and pressures, ADAFs inflate to become geometrically ``thick" and sub-Keplerian. Another interesting aspect of disks in this accretion regime is the possibility of generating significant outflows \citep{adios,gonewiththewind}.  

The inability to radiate enough energy also arises for very high accretion rates, giving rise to advection dominated flows once the disk becomes radiation pressure dominated \citep{Paczynsky1980,Jaro:1980}.

The energy release rate in a Keplerian disk, down to a radius $r$, is given by $GM_\mr{BH}\dot{m}/2r$, where $M_\mr{BH}$ is the black hole mass, and $\dot{m}$ is the mass accretion rate. This energy should be compared to the Eddington luminosity of the BH, defined as:
\begin{equation}
L_\mr{Edd}\equiv \frac{4\pi cGM_\mr{BH}}{\kappa},
\label{eq:ledd}
\end{equation}
from which a critical accretion rate can be define to be
\begin{equation}
\dot{m}_\mr{crit}\equiv\frac{L_\mr{Edd}}{\eta_0 c^2},
\label{eq:mdotcrit}
\end{equation}  
with $\eta_0=1/16$ being the standard efficiency for accretion around a Schwarzschild BH. If $\dot{m}$ is large enough, that is, $\dot{m} \gtrsim \dot{m}_\mr{crit}$ then the energy released will approach the Eddington luminosity before reaching the inner radius of the disk. From that point inwards, the radiation pressure will be dominant. The disk will become geometrically thick (having a scale height $H(r) \sim r$), and the local flux will approach the Eddington flux defined as
\begin{equation}
F_\mr{Edd}\equiv\frac{cGM_\mr{BH}z}{\kappa R^3},
\label{eq:Fedd}
\end{equation}
with $R \equiv \sqrt{r^2 + z^2}$ (where $r$ and $z$ are the cylindrical coordinates). 
However, because the Eddington flux cannot be surpassed,  the dissipated energy cannot be entirely radiated away and part of it should therefore be advected inwards. The high pressure also implies that the disk rotates at sub-Keplerian velocities.

Other interesting aspects of these disks is their possible ability to accelerate very luminous jets \citep{AP1980}, and the fact that although the local flux in the disk does not surpass the Eddington flux, the overall luminosity can surpass the Eddington luminosity, simply due to the disk geometry \citep{Jaro:1980, Paczynsky1980}.  Roughly, the Eddington luminosity can be surpassed by a factor of $\sim \ln r_{out}/r_{in}$, where the radii denote the inner and outer extents of the radiation pressure disk. 

For nearly critical accretion, an intermediate type of solutions exists, that of ``slim-disks". However, because it exists only near the last stable orbit (thus allowing for advection of heat into the Roche Lobe overflow), its solution requires relativistic corrections \citep{Abramowicz1988}. 

The above models assume, however, that the Eddington flux cannot be surpassed locally. Nevertheless, it was shown that super-Eddinigton states do naturally arise in nature \citep{ShavivNovae}, allowing for high luminosities, while generating optically thick winds. Our goal in the present work is to consider the recent advances in the understanding of how super-Eddington atmospheres arise, and what they look like, and to incorporate these ideas into models for very high accretion rate accretion disks.

In \S\ref{sec:background}, we begin by reviewing our present understanding of how super-Eddington states arise. In \S\ref{sec:model} we describe our model for super-Eddington accretion, and in \S\ref{sec:numerical} we describe the numerical solution. In \S\ref{sec:results} we describe the numerical results, and end with a discussion in \S\ref{sec:discussion}. 

\section{Background: Super-Eddington States}
\label{sec:background}

We begin by reviewing the relevant physics pertaining to the emergence of super-Eddington states. These include three particular elements. First, the rise of inhomogeneities due to radiative-hydrodynamic instabilities was shown to reduce the effective opacity \citep{ShavivPorous}. This allows for the existence of super-Eddington atmospheres. Second, once a super-Eddington state arises, strong continuum driven winds are accelerated \citep{ShavivEta,ShavivNovae,OGS}. These optically thick winds are responsible for a significant mass loss and are also important when determining the appearance of these objects. Last, if the wind mass loss is too large, wind stagnation and a photon-tired state arises \citep{Owocki,vanMarle}. In it, a layer is formed in which strong shocks mediate a high energy flux without an excessive  mass flux. These three components are the necessary building blocks for the Super-Eddington (SED) accretion disk models, and we therefore review them below. 

\subsection{The rise of super-Eddington states}
\label{sec:SEDstate}

According to common wisdom, objects cannot shine beyond their classical
Eddington limit, $\ledd$ (or locally beyond the Eddington flux, $\fedd$, as is the case in accretion disks), since no hydrostatic solution exists.  In
other words, if objects do pass $\fedd$, they are highly dynamic.
They have no steady state, and a huge mass loss should occur since their
atmospheres are then gravitationally unbound and they should therefore
be expelled.  Thus, astrophysical objects according to this picture, can
pass $\ledd$ but only for a short duration corresponding to the time
it takes them to dynamically stabilize once SED conditions arise.

For example, this can be  seen in detailed 1D
numerical simulations of thermonuclear runaways in classical nova eruptions, which can achieve SED luminosities
but only for several dynamical time scales \citep[e.g.,][]{Starrfield1989}.
However, once they do stabilize, they are expected and indeed do reach
 in the simulations, a sub-Eddington state.  Namely, we
naively expect to find no steady state SED
atmospheres. This, however, is not the case in nature, where nova eruptions are clearly SED for durations which are orders of magnitude longer then their dynamical time scale \citep{ShavivNovae}. This is exemplified with another clear SED object---the great eruption of the massive star $\eta$-Carinae, which was a few times above Eddington for over 20 years \citep{ShavivEta}

The existence of a super-Eddington state can be naturally explained, once we consider the following:
\begin{enumerate}
\item Atmospheres become unstable as they approach the Eddington limit. In addition to instabilities that operate under various special conditions (e.g., Photon bubbles in strong magnetic fields, \citealt{Bubbles1,Bubbles2,Bubbles3}, or s-mode instability under special opacity laws, \citealt{Opacity1,Opacity2}), two instabilities operate in Thomson scattering atmospheres \citep{ShavivInstabilities}. It implies that {\em all atmospheres will become unstable already before reaching the Eddington limit}.
\item The effective opacity for calculating the radiative force on an inhomogeneous atmosphere is not necessarily the microscopic opacity. Instead, it is given by
\begin{equation}
\label{eq:effectiveOpacity}
    \kappa_{V}^{\mr{eff}} \equiv {\left\langle F\kappa_{V}\right\rangle_{V}
      \over  \left\langle F \right\rangle_{V}},  
\end{equation}
where $\left< ~\right>_V$ denotes volume averaging and $F$ is the flux \citep{ShavivPorous}.
 The situation is very similar to the Rosseland vs. Force opacity means used in non-gray atmospheres, where the inhomogeneities are in frequency space as opposed to real space. For the special case of Thomson scattering, the effective opacity is always reduced.

\end{enumerate}

Thus, we find that as atmospheres approach their classical Eddington limit, they will necessarily become inhomogeneous. These inhomogeneities will necessarily reduce their effective opacity such that the effective Eddington limit will not be surpassed even though the luminosity can be super-classical-Eddington. This takes place in the external regions of luminous objects, where the radiation diffusion time scale is shorter than the dynamical time scale in the atmosphere. Further inside the atmosphere, convection is necessarily excited such that the total energy flux may be SED, but the radiative part of it is necessarily sub-Eddington with the convective flux carrying the excess \citep{Joss1973}. 

\subsection{Super-Eddington Winds}
\label{sec:SEDwinds}
The atmospheres of SED objects, and SED accretion disks in particular, effectively remain sub-Eddington while being classically super-Eddington, only as long as the inhomogeneities comprising them are optically thick. This condition will break at some point where the density is low enough. At this height, the effective opacity returns to its microscopic value and hence the radiative force becomes super-Eddington again. From this point outwards we obtain continuum driven winds. Because the winds are generally optically thick, the conditions in them affect the structure of the disk beneath. 

At the critical point, the radiative and gravitational forces balance each other. This point will coincide with a sonic surface for a steady state wind (where the mass loss velocity equals the local speed of sound). This allows us to obtain the local mass loss rate per unit area, given by
\begin{equation}
\dot{\phi}_{\mr{wind}}=\rho_{\mr{crit}}v_s({z_{\mr{atm}}})=\mathit{const.},
\label{eq:phidot}
\end{equation}
where $z_{\mr{atm}}$ is the vertical height of the critical point, $\rho_{\mr{crit}}$ is the density at this point and $v_s$ is the local speed of sound.

Based on the fact that instabilities develop structure with a typical size comparable to the density scale height in the atmosphere, it is possible to estimate the average density at the sonic point \citep{ShavivNovae}. Using this density, the mass loss can be estimated to be
\begin{equation}
\dot{\phi}_{\mr{wind}}={\cal W}\frac{F-\fedd}{cv_s}.
\label{eq:phidot2}
\end{equation}
where $\func$ is a dimensionless wind ``function''.  In principle, $\func$
can be calculated ab initio only after the nonlinear state
of the inhomogeneities is understood.  This however is still lacking
as it requires elaborate 3D numerical simulations of the nonlinear
steady state.  

Nevertheless, deriving $\func$ can be achieved in several
phenomenological models which depend on geometrical parameters such as the average size of the inhomogeneities in units of the
scale height ($\zeta \equiv d / l_{p}$), the average ratio between the surface area and volume of
the blobs in units of the blob size ($\Xi$), and the volume filling
factor $\eff$ of the dense blobs.
For example, in the limit in which the blobs
are optically thick, one can show that
$
\func \approx {3 \Xi / 32 \sqrt{\nu} \eff \zeta (1-\eff)^{2}}
$ \citep{ShavivNovae},
with $\nu$ being the ratio between the effective speed of sound in the
atmosphere to the adiabatic one. Thus, $\func$ depends only on
geometrical factors. It does not depend explicitly on the Eddington
parameter $\Gamma \equiv L/\ledd $ as long as the blobs have a single length scale. Once this assumption is alleviated, $\func$ can become a weak function of $\Gamma$ \citep{OGS}. Comparison to observations yields typical values of $\func \sim 1-10$ \citep{ShavivNovae}.

\subsection{Photon-Tired Winds}
\label{sec:photon-tired}

An interesting modification to the above continuum driven winds arises when the predicted mass loss is too high for the available luminosity to push it to $r\rightarrow \infty$. This happens when $v_{esc} \gtrsim \sqrt{ v_s c / {\cal W}}$, and it gives rise to ``photon-tired winds" \citep{Owocki}.  A wind solution with a monotonically decreasing velocity is then not possible, because the wind stagnates at a finite radius.

The behaviour of photon tired winds was studied by \cite{vanMarle}. It was found that shocks form between infalling material and the outflowing wind. This forms a  layer of shocks in which there is a large kinetic flux, but without the associated mass flux. When photon tired winds arise, the mass loss from the top of the layer of shocks is reduced to less than the photon tired limit, and the luminosity to less than the Eddington luminosity.

\section{The Model}
\label{sec:model}

\begin{figure*}
\centerline{\epsfig{file=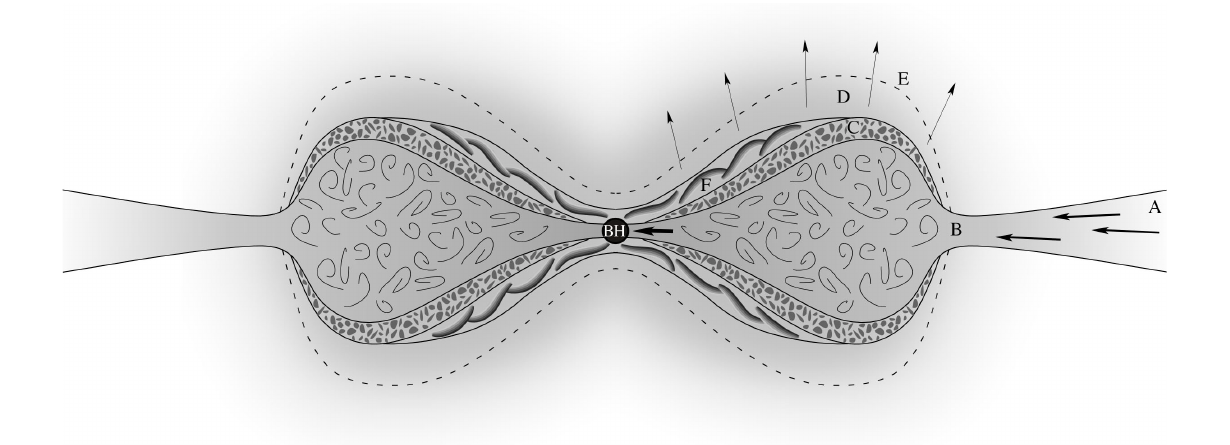,width=16cm}}
\caption{The model structure. The different regions are: (A) A sub-Eddington thin disk (following the solution of S\&S). (B) The accretion disk 
loses its flat geometry at the radius where the energy release corresponds to the Eddington luminosity. The disk inflates and becomes radiation-pressure dominated. Once super-Eddington states 
are allowed to arise, the standard thick disk picture is modified. (C) A porous layer forms at the less dense regions above 
the convection layer. (D) At a height were the porous structure become optically thin, a wind is accelerated. (E) 
Since it is optically thick, the photosphere is located in the wind itself. (F) In the inner parts of the disk, the 
escape velocity is large enough to give rise to a photon tired layer, which effectively moves the sonic point higher. 
 }
\label{fig:themodel}
\end{figure*}

Our goal is to construct Super-Eddington accretion disks, namely, disks which radiate fluxes that can locally exceed the Eddington flux. We look for slim disk solutions which are heuristically described in fig.\ \ref{fig:themodel}. More specifically, we have to consider the following points. 

\begin{itemize}
\item {\em Geometry:} Although in principle, one could envision super-Eddington accretion solutions with different geometries, we look for disk like solutions. We shall assume that the vertical length scale is sufficiently smaller than the radius, such that we can deconstruct the problem into vertical and radial components. As we shall see below, the hydrostatic part of the disk satisfies this conditions except for the highest accretion rates. Note that because the vertical structure is not {\em much} smaller, the geometry is not that of a thin disk, but that of a slim disk. This also implies that we cannot assume Keplerian velocities. 
\item {\em Porosity:} As mentioned above, high radiative fluxes give rise to porosity which reduces the effective opacity, thereby allowing the existence of super-Eddington fluxes in the hydrostatic atmospheres. Thus, a necessary component of our model is an opacity law of the form $\kappa(\Gamma)$, which takes the porosity into account.  
\item {\em Convection:} \cite{Joss1973} have shown that high radiative fluxes give rise to convection as the radiative fluxes approach the Eddington limit. The dense inner parts of the SED disk, near the equatorial plane are therefore expected to be convective. 
\item {\em Wind:} Since a porous atmosphere can reduce the opacity only as long as the inhomogeneities comprising it are optically thick, a necessary outcome of SED atmospheres is the acceleration of continuum driven winds where the average density is low enough. Because this mass loss can be significant, it has two interesting ramifications. First, because the wind is generally optically thick, the photosphere is going to reside in the wind. This will have various observational consequences. Second, the mass accretion rate decreases as the radius decreases. 
\end{itemize}

\subsection{Radial Structure}
As mentioned above, the small thickness of the disk enables the separation between the radial structure and the vertical one.  The equations describing the radial structure are obtained from the radial conservation of mass, radial momentum, energy, and angular momentum. An additional equation is the closure relation for the stress tensor. 

Radial mass conservation gives:
\begin{equation}
\frac{d\dot{\mr{m}}}{dr}=4{\pi}r\dot{\phi}_\mr{wind},
\label{eq:dmdot}
\end{equation}
where
\begin{equation}
\dot{\mr{m}}=4{\pi}rv_r\int_0^H{\rho}dz
\label{eq:mdot}
\end{equation}
is the mass accretion rate. Note that we assume a height independent velocity structure.

Conservation of radial momentum gives
\begin{equation}
v_r\frac{dv_r}{dr}+\frac{1}{\rho}\frac{dP}{dr}= -{\partial \Psi \over \partial r},
\label{eq:radial_momentum}
\end{equation}
where $\Psi$ is a pseudo-Newtonian potential given by \citet{Paczynsky1980} as
\begin{equation}
\Psi=-\frac{GM_{\mr{BH}}}{R-rg},
\label{eq:psi}
\end{equation}
and $R=\sqrt{r^2+z^2}$. For the radial structure, we assume that $z=0$. 

The equation for angular momentum conservation is
\begin{equation}
{\rho}v_r\frac{d}{dr}\left(r^2\omega\right)=-\frac{1}{r}\frac{d}{dr}\left(r^2\tau_{r\phi}\right),
\label{eq:angular_momentum}
\end{equation}
where $\tau _{r\phi}$ is the tangential stress. Following the standard ``$\alpha$" prescription, we write:
\begin{equation}
\tau_{r\phi}=-{\alpha}P.
\label{eq:stress_tensor}
\end{equation}

The heat produced by the viscosity is partly radiated away (locally), and partly transferred by advection into smaller radii. The radial heat advection is given by the difference between the amount of heat being produced by the viscosity, and the energy radiated away from the surface of the disk, that is,
\begin{equation}
T\frac{ds}{dr}=\left(F-\Phi\right)\frac{4{\pi}r}{\dot{\mr{m}}},
\label{eq:advection}
\end{equation}
where $s$ is the specific entropy, $F$ is the flux radiated from the surface, while $\Phi$ is the dissipation function, given by
\begin{equation}
\Phi=r\frac{d\omega}{dr}\int_0^{z_0}\tau_{r\phi}dz.
\label{eq:dissipation}
\end{equation}
Note that the specific entropy increases with decreasing radii in the case of advection, hence $ds/dr<0$. However, another possibility exists, in which the infalling matter releases part of its heat (i.e., $ds/dr>0$). This heat, together with the heat generated by the viscosity, is radiated away. This process dominates the inner radii region of the accretion flow, as we will show in the results.

\subsubsection{Inner Sonic Point}
At large radii, the radial velocity is very small when compared to the speed of sound, while matter is freely infalling at the vicinity of the BH. This implies that at some radius $r=r_s$, the radial velocity should become equal to the speed of sound, and then exceed it. Similarly to the case of spherical Bondi accretion, we can obtain the condition for this point. Because of the radial pressure gradient, this point will reside between the last stable orbit $r_{ls}=3r_g$ and the marginally stable orbit, $r_{ms}=2r_g$. Using eqs.\ \ref{eq:mdot}-\ref{eq:radial_momentum}, and that
\begin{equation}
\frac{dp}{dr}=\frac{\partial{p}}{\partial{s}}\frac{ds}{dr}+\frac{\partial{p}}{\partial{\rho}}\frac{d\rho}{dr}=\frac{p}{s}\left(\frac{4}{3}-\beta\right)\frac{ds}{dr}+v_s^2\frac{d\rho}{dr},
\label{eq:dp}
\end{equation}
where $v_s^2\equiv\partial{p}/\partial{\rho}$, and $s$ is the specific entropy, we obtain 
\begin{equation}
\frac{v_r^2-v_s^2}{v_r}\frac{dv_r}{dr}\simeq\frac{1}{r}\left[v_s^2+(v_{\phi}^2-v_\mr{kep}^2)\right]-\frac{\partial{p}}{\partial{s}}\frac{ds}{dr}{1 \over \rho}.
\label{eq:sonic1}
\end{equation}
Note that we have assumed the following approximation, that
\begin{equation}
\frac{1}{\Sigma}\frac{d\Sigma}{dr}\approx \frac{1}{\rho}\frac{d\rho}{dr}
\label{eq:sigma}
\end{equation}
where $\Sigma\equiv\int\rho dz$.
At the sonic radius $v_r=v_s$, such that the r.h.s. of eq.\ \ref{eq:sonic1} must vanish, i.e., at the sonic radius we have that
\begin{equation}
v_s^2+(v_{\phi}^2-v_\mr{kep}^2) \simeq r\frac{\partial{p}}{\partial{s}}\frac{ds}{dr}\frac{1}{\rho}.
\label{eq:sonic2}
\end{equation}
We use this expression as the inner boundary condition for the disk. 

\subsection{Vertical Structure}

The vertical structure of the SED accretion disk can be divided into two regions, a hydrostatic region which includes also the porous atmosphere, and the region of a continuum driven wind. Because the wind is optically thick, the thermal conditions at the wind affect the hydrostatic structure. This is unlike typical stellar systems with optically thin winds. As a consequence, the hydrostatic structure has to be solved together with the wind, though the governing equations are different for the two regions. 

\subsubsection{Hydrostatic Region}
In the hydrostatic region, any energy which is generated by the viscosity and not advected radially, is transported in the vertical direction through either convection or radiative transfer, or both.  

The first equation describing this region is that of hydrostatic equilibrium,
\begin{equation}
\frac{1}{\rho}\frac{dP}{dz}=-\frac{d\Psi}{dz}.
\label{eq:hydro_eq_z}
\end{equation}

The temperature gradient is determined according to the energy transfer mechanism. It is given by
\begin{equation}
\frac{dT}{dz}=\left\{\begin{array}{ll} \displaystyle {\gamma-1 \over \gamma}\frac{dP}{dz}\frac{T}{P}, & \textrm{in the convective zone,}\\
\\
\displaystyle -\frac{3{\kappa_{\mr{eff}}}{\rho}F}{4acT^3}, & \textrm{in the radiative zone},\\
\end{array}\right.
\label{eq:gradT}
\end{equation}
where $\rho$ is the density, $F$ is the vertical radiative flux and $\gamma$ is the adiabatic index. Convection is present if the standard Schwarzschild criterion is satisfied. But for convection to be efficient, the convective flux must be smaller than the maximum possible which is given by
\begin{equation}
F_\mr{conv,max} =\rho v_s^3,
\end{equation}
and $v_s$ is the adiabatic speed of sound.
The opacity in the radiative zone is taken to be the Thomson opacity, as long as the radiative flux is smaller than the critical flux above which the atmosphere develops inhomogeneities. As described in \S\ref{sec:SEDstate}, the gas becomes inhomogeneous above the critical flux, such that the radiative force exerted on the gas is reduced. 

We assume that the relation between the effective Eddington factor $\Gamma_{\mr{eff}} \equiv F/\feff$ and the classical Eddington factor $\Gamma \equiv F/\fedd$ is empirically given by
\begin{eqnarray}
\Gamma_{\mr{eff}} & = & 1-\frac{A}{\Gamma^B}~~{\mr{for}}~~\Gamma > \Gamma_{\mr{crit}},  \nonumber \\
\Gamma_{\mr{eff}} & = & \Gamma~~{\mr{for}}~~\Gamma < \Gamma_\mr{crit}.
\end{eqnarray}
$\Gamma_{\mr{crit}}$ is the critical $\Gamma$ above which inhomogeneities are excited, so the effective opacity for $\Gamma>\Gamma_{\mr{crit}}$ is given by
\begin{equation}
\kappa_{\mr{eff}}=\kappa_{\mr{Th}}\left(1-\frac{A}{\Gamma^B}\right)/\Gamma.
\label{eq:kappa_eff}
\end{equation}
Since we expect a continuous $\Gamma_{\mr{eff}}$, $A$, $B$ and $\Gamma_{\mr{crit}}$ satisfy the equation $\Gamma_{\mr{crit}} = 1- A / \Gamma_{\mr{crit}}^B$. 
From theoretical considerations, we take $\Gamma_{\mr{crit}} \sim 0.8$ \citep{ShavivInstabilities}. This implies a relation between the normalization constant $A$ and the power law $B$, that is given by
\begin{equation}
A=(1-\Gamma_{\mr{crit}}) \Gamma_{\mr{crit}}^B.
\label{eq:A}
\end{equation}
Note that because the behaviour of super-Eddington atmospheres is expected to depend on only the local conditions, the parameters are taken to be location independent. 

\subsubsection{Continuum Driven Winds}
As described in \S\ref{sec:SEDwinds}, a continuum driven wind is accelerated from the region where the density is low enough, such that the inhomogeneity based opacity reduction becomes inefficient. In this region, the effective opacity approaches the microscopic value, such that the radiative flux is again super-Eddington. 

The primary equations describing the wind structure are the equation of motion
\begin{equation}
{\rho}v_z\frac{dv_z}{dz}=-\frac{dP}{dz}-{\rho}g_z,
\label{eq:wind_motion}
\end{equation}
where $g_z=-\partial\Psi/{\partial}z$, and the energy conservation equation
\begin{equation}
F(z)=F_\mr{atm}-\dot{\phi}_\mr{wind}\left(\frac{v_z^2}{2}+\frac{GM_\mr{BH}}{R_\mr{atm}}-\frac{GM_\mr{BH}}{R}\right).
\label{eq:energy_conservation}
\end{equation}
The index ``$\mr{atm}$" denote values at the top of the hydrostatic atmosphere beneath the wind.
Note that we assume here that the wind geometry is that of a slab. Namely, we assume that $z \ll R$. This assumption breaks down for high accretion rates, at which point the present solution fails. 

From the last two equations, and the assumption that $\kappa=const.$, we derive the radiative flux and wind velocity as a function of $z$:
\begin{equation}
F(z)=F_\mr{atm} \exp\left(\frac{\kappa\dot{\phi}_\mr{wind}(z_\mr{atm}-z)}{c}\right),
\label{eq:Fwind}
\end{equation}
\begin{equation}
\frac{v_z^2}{2}=\frac{GM_{BH}}{R_\mr{atm}}\left[\frac{1}{m}\left(1-\frac{F(z)}{F_\mr{atm}}\right)+\left(\frac{R_\mr{atm}}{R}-1\right)\right]+\frac{v_s^2}{2},
\label{eq:vz2}
\end{equation}
where 
\begin{equation}
m\equiv\frac{\dot{\phi}_\mr{wind}GM_{BH}/R_\mr{atm}}{F_\mr{atm}}.
\label{eq:m}
\end{equation}
$m$ is the ratio between the energy flux needed to accelerate the wind out of the gravitational potential well, and the radiation flux provided to the wind by the system.

Another aspect of this thick wind is the location of the photosphere. While in the slim and thin disk models the photosphere resides where the gas becomes optically thin, in our case, the photosphere resides much higher, where the wind becomes optically thin. The optical depth of the wind is given by
\begin{equation}
\tau=\int_{z_\mr{atm}}^\infty\kappa{\rho}dz.
\label{eq:tau}
\end{equation}

Note also that the change in the location of the photosphere is accompanied by a decrease in the radiative flux emitted from the disk (as energy is used to accelerate the wind), hence, a decrease in the effective temperature.

As described in \S\ref{sec:results}, the typical ratio we obtain between the height of the photosphere and the radius is small, that is, $z_{ph}(r)/r \lesssim 1$, as long as the accretion rates are not too large, i.e., $\dot{\mr{m}} \lesssim 20\dot{\mr{m}}_{\mr{crit}}$. For higher accretion rates, the wind geometry ceases to be disk-like, and the solution described here breaks down. 

\subsubsection{Photon Tired Winds}
As elaborated upon in \S\ref{sec:photon-tired}, photon tired winds are formed when the available radiative flux at the sonic point is insufficient to drive the mass loss driven at the sonic point out of the gravitational potential well  \citep[see][]{Owocki}.
Under such conditions, a layer of shocks forms in which the effective sonic point moves upwards, and reduces the actual mass loss. Using the results from \cite{vanMarle}, we empirically model the maximal $\dot{\phi}_\mr{wind}$ to be,
\begin{equation}
\frac{\dot{\phi}_\mr{wind}}{\dot{\phi}_\mr{tiring}} \simeq max\left(0.2\left(\frac{F}{F_\mr{Edd}}\right)^{0.6}, 0.9\right),
\label{eq:max_phidot}
\end{equation}
where $\dot{\phi}_\mr{tiring}{\equiv}F/(GM_\mr{BH}/R_\mr{atm})$.

\subsubsection{Boundary Conditions}
The vertical structure of the disk is determined by the local equatorial conditions, i.e., the total pressure $P(r,z=0)$ and the density $\rho(r,z=0)$.  The total radiative flux is a free variable. It is determined by the vertical structure which has to be solved for given the top boundary condition for the radiation field. The latter is the blackbody radiation law,
\begin{equation}
F(r,z_0)={\sigma}T_\mr{eff}^4.
\label{eq:bb_condition}
\end{equation}
Here $T_\mr{eff}$ is the temperature at optical depth $\tau=2/3$, as obtained by the outward integration.

\section{Numerical Solution}
\label{sec:numerical}
The problem we are required to solve is devided into two parts, vertical and radial integration. Each radial integration step  is accompanied by a vertical integration. 

For the vertical integration we take an initial guess for the radiative flux and solve eqs.\ \ref{eq:hydro_eq_z}-\ref{eq:gradT} for the hydrostatic part, and eqs.\ \ref{eq:wind_motion}-\ref{eq:energy_conservation} for the wind (equations are integrated using the $4^\mr{th}$ order Runge-Kutta method), and the validity of the boundary condition (eq. \ref{eq:bb_condition}) is checked. This vertical integration is repeated with different values for the radiative flux (using the shooting method) until eq. \ref{eq:bb_condition} is fulfilled. 

Using the emitted flux $F$, the mass loss rate $4\pi r\dot{\phi}_\mr{wind}dr$ and the vertical integrations over the pressure  and the density, eqs.\ \ref{eq:dmdot}-\ref{eq:radial_momentum}, \ref{eq:angular_momentum} and \ref{eq:advection} are solved. This integration is taken up to the inner sonic radius (where $v_r=v_s$), which should reside between $r_{ls}$ and $r_{mb}$. The location of the sonic point is determined by the initial guess for the outer pressure, $p(r_\mr{out},0)$. A shooting method is used  to determined the outer pressure.

\section{Results}
\label{sec:results}

\begin{figure}
\centerline{\epsfig{file=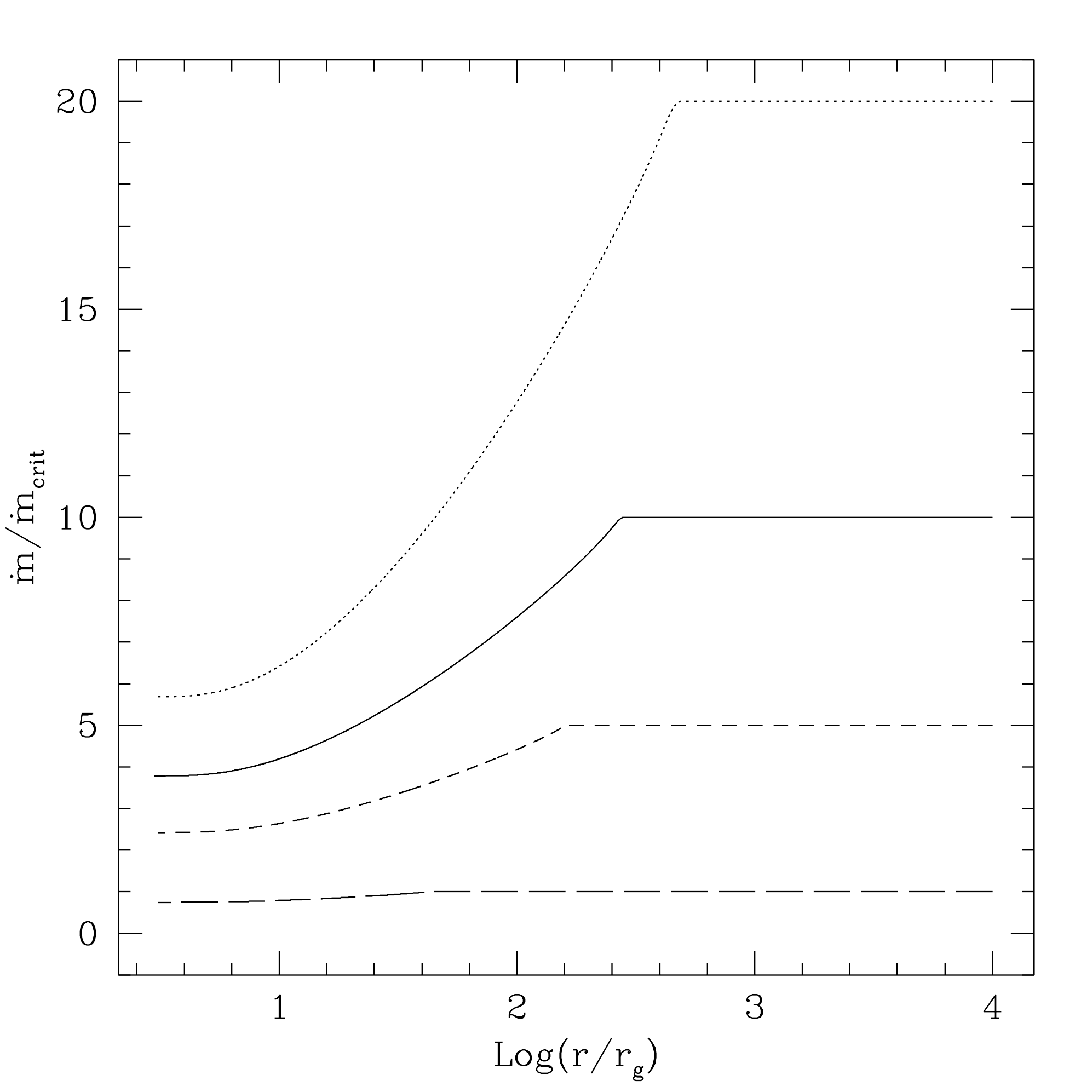,width=8cm,angle=0}}
\caption{The mass accretion rate (in units of the critical rate, $\dot{m}/\dot{m}_{crit}$) as a function of radius, for different outer accretion rates onto a BH with $M_{bh}=10M_{\odot}$. The dotted line denotes $20\dot{m}_{crit}$, the solid line denotes $10\dot{m}_{crit}$, the short dashed describes the $5\dot{m}_{crit}$ case, and long dashed line accretion with $\dot{m}_{crit}$.}
\label{fig:mdot}
\end{figure}

\begin{figure}
\centerline{\epsfig{file=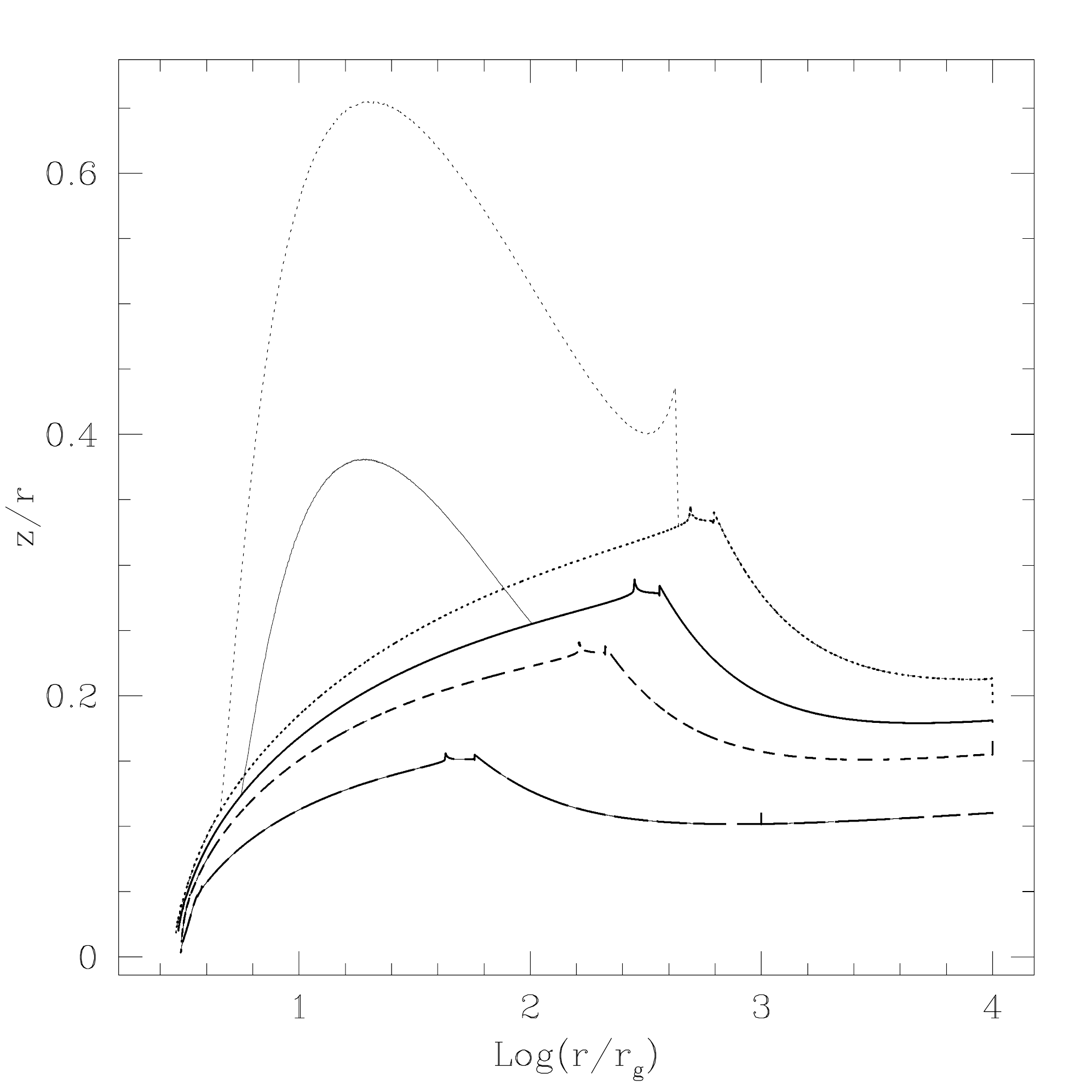,width=8cm,angle=0}}
\caption{The vertical height ($z/r$) of the disk as a function of radius, for the same accretion rates as before. The thick lines denote the position of the photosphere when a wind is absent, while the thin lines denote the photosphere in the thick wind when present, in which case the thick line denotes the location of the sonic surface. The line types are as before. Note that for the two lower accretion rate $\dot{m}=\dot{m}_{crit}$ and $\dot{m}=5\dot{m}_{crit}$ the wind is always optically thin and the photosphere coincides with the sonic point of the wind for all radii.}
\label{fig:z}
\end{figure}

\begin{figure}
\centerline{\epsfig{file=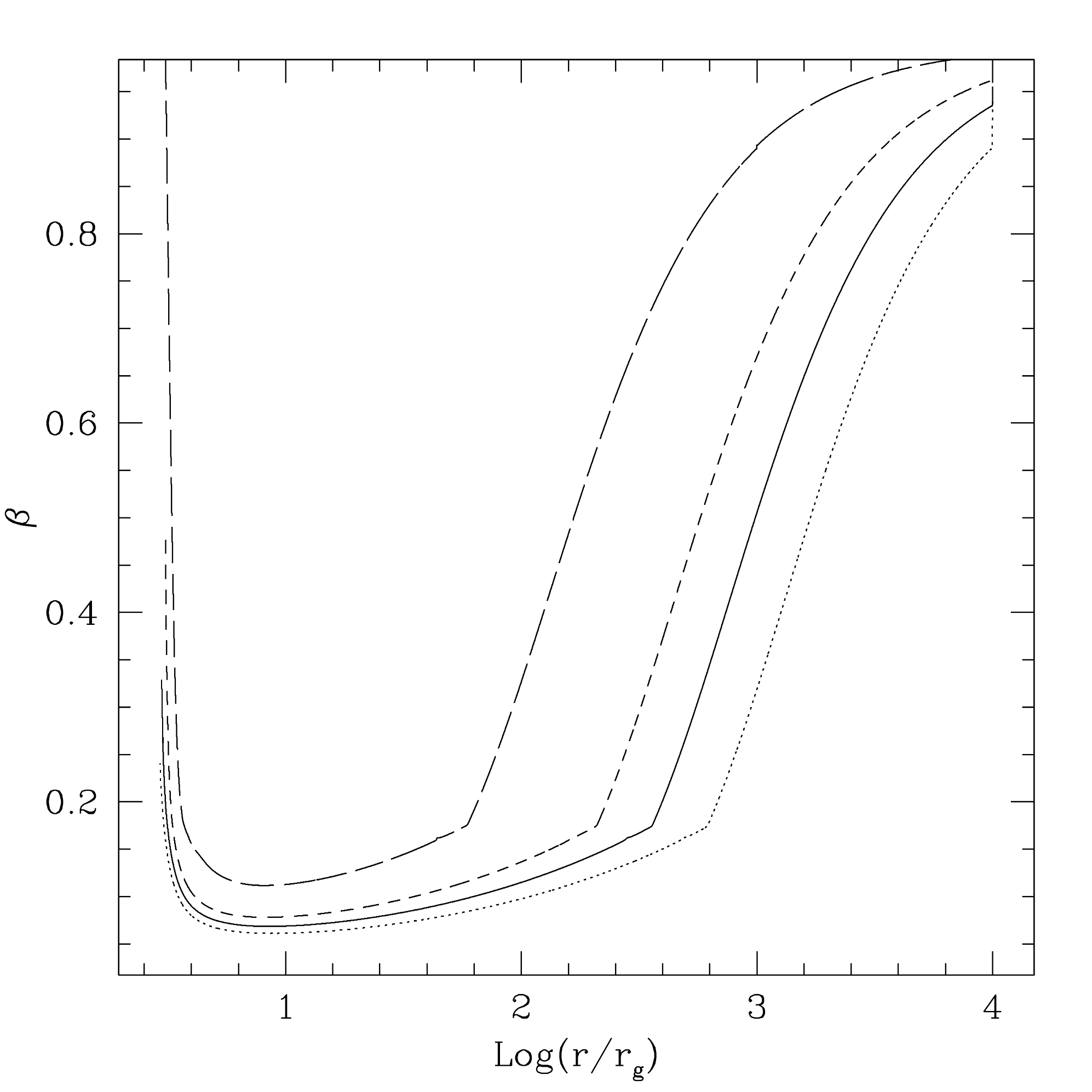,width=8cm,angle=0}}
\caption{The ratio $\beta{\equiv}P_g/P_{tot}$ vs. $\log(r/r_g)$. At large radii, the solutions approach the standard solution, and the radiation pressure is less important. Closer to the black hole, the radiation pressure becomes more important and the radiative flux increases. Near the inner radial sonic point $ds/dr>0$, and the gas pressure becomes important again.}
\label{fig:beta}
\end{figure}

In this section we present the results for accretion disks with $\alpha=0.001$ around a stellar BH of mass $10M_{\odot}$ and around a super-massive BH of mass $10^{6.5}M_{\odot}$. We define the {\em outer accretion rate} as the rate of mass entering the disk at large radii before any mass loss takes place.

 For the stellar BHs, we take outer accretion rates of  1, 5, 10 and 20 $\dot{m}_\mr{crit}$. We find that the disks lose considerable amounts of mass through a wind, such that the real accretion rates (= mass passing the sonic radius per unit time, $\dot{m}_\mr{real}$) are 0.74, 2.4, 3.8, and 5.7 respectively, i.e, in stellar BHs with outer accretion rates $\mr{\dot{m}}\gtrsim 5\mr{\dot{m}_\mr{crit}}$ more than half of the mass entering the accretion disk in the outer radius will not be accreted into the black hole but leave as a wind. Mass accretion rates as a function of the radius are shown in fig.\ \ref{fig:mdot}.
The vertical heights of the disks are given in fig.\ \ref{fig:z}, where thin lines denote the location of the photosphere inside the wind, while the thick lines denote the sonic surface (the base of the wind). At large radii, where the wind is absent, thick lines give the heights of the photosphere as well. Note that it is impossible to obtain a situation in which all the mass accelerates in a wind since part of the energy is used to heat the gas, implying that without any net accretion, there will be an insufficient amount of energy to drive the matter back to infinity.

The gas to total pressure ratio, $\beta$ in the equatorial plane ($z=0$), is shown in fig.\ \ref{fig:beta}. Apparently, the gas pressure dominates at large radii. However, radiation pressure becomes progressively more important at smaller radii. Radiation pressure is also dominant for higher accretion rates. Close to the radial sonic point the gas pressure becomes important again.
\subsubsection{Efficiency of Accretion}
The efficiency of accretion is defined by
\begin{equation}
\eta=\frac{L}{\dot{m}c^2}
\label{eq:eta}
\end{equation}
Substituting $\dot{m}_{real}$ into (\ref{eq:eta}) we obtain the accretion efficiencies (see table \ref{tab:results}). For low accretion rates, the efficiency is slightly higher than the pseudo-Newtonian $\eta_0$ case because the wind is optically thin, while for higher accretion rates the wind is optically thick and $\eta/\eta_0<1$.

\subsubsection{Spectra and Luminosities}
The total luminosities leaving the disks in units of Eddington luminosity are 0.8, 2.65, 3.9 and 4.85  respectively. Thus, for outer accretion rates of about 1.5 $\dot{m}_\mr{crit}$, the total radiated luminosity exceeds the Eddington ``limit". 

\begin{figure}
\centerline{\epsfig{file=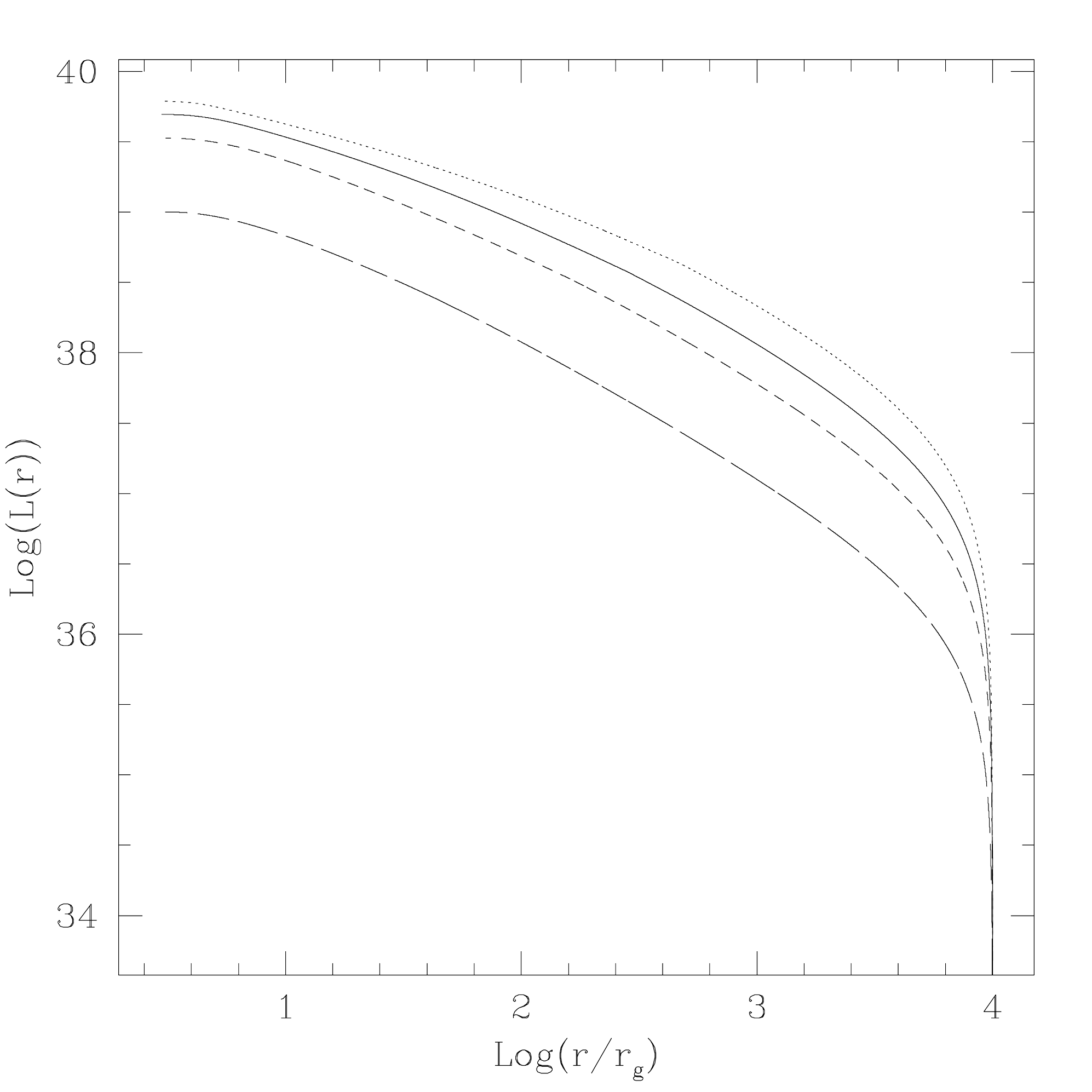,width=8cm,angle=0}}
\caption{Radially integrated luminosity $\log(L(r))$ vs. $Log(r/r_g)$ for the same disks as before.}
\label{fig:Lr}
\end{figure}

\begin{figure}
\centerline{\epsfig{file=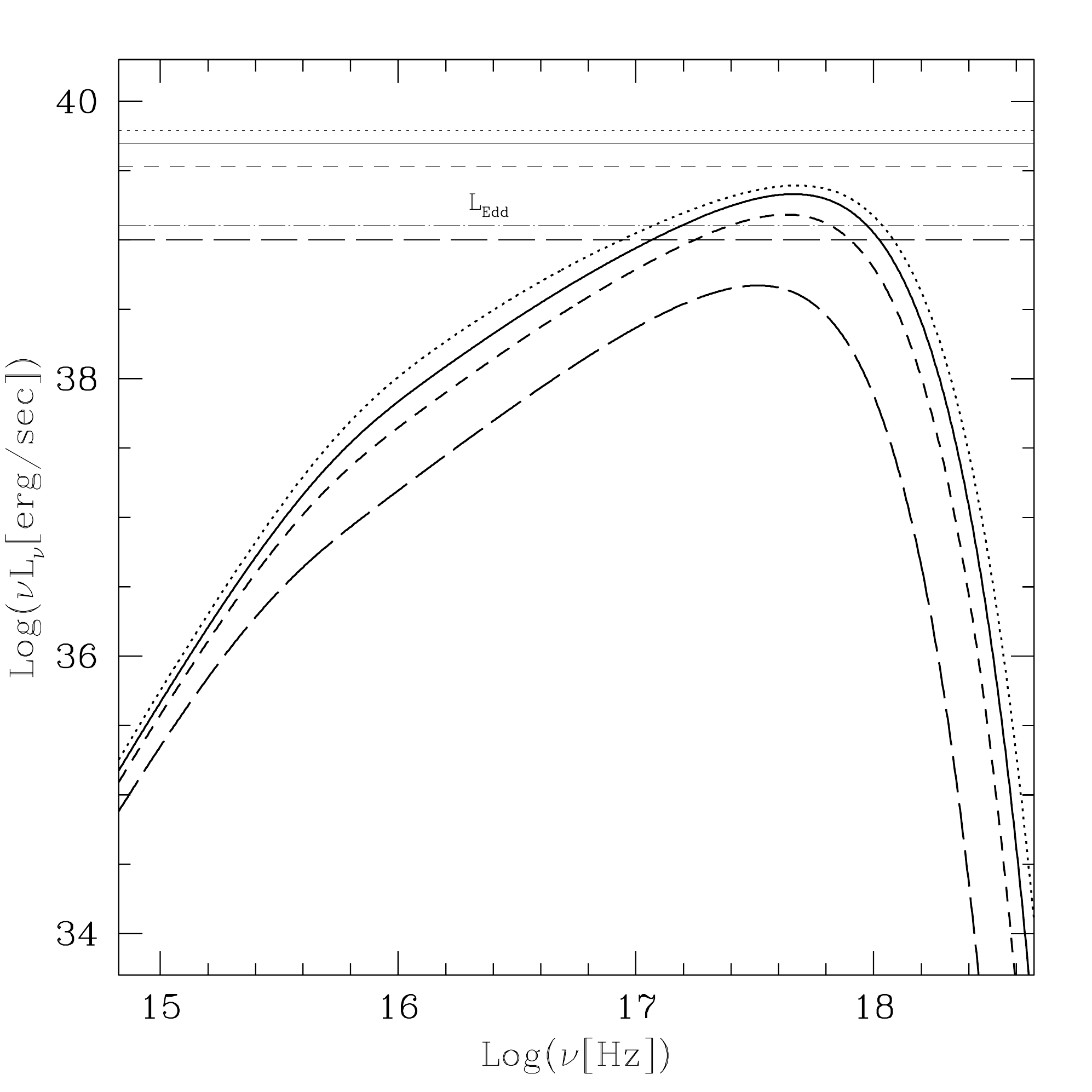,width=8cm,angle=0}}
\caption{The total luminosity $\log({\nu}L_{\nu})$ vs. $\log(\nu)$, for the same disks as before. }
\label{fig:L}
\end{figure}

The accumulated luminosity $L(r)=\int_{r_{out}}^r4{\pi}rF(r)dr$ is shown in fig.\ \ref{fig:Lr}. Assuming a local Planck distribution and using the effective temperature $T_\mr{eff}(r)$, the emergent luminosity per unit frequency, $L_{\nu}$, is
\begin{equation}
L_{\nu}={\pi}\int_{r_s}^{r_\mr{out}}4{\pi}rB_{\nu}(T_\mr{eff}(r))dr
\label{eq:Lnu}
\end{equation}
where $B_{\nu}$ is Planck function.
This is given in fig.\ \ref{fig:L}. 

Table \ref{tab:results} summarizes the above results.

\subsubsection{Sensitivity to Model Parameters}
There are two main uncertainties in our model. The first is in the basis of all $\alpha$-disk models, the $\alpha$ parameter. The only theoretical limit on it is that $\alpha<1$. We checked the sensitivity of our model using three different values of $\alpha$, $\alpha=0.001,0.01$ and $0.1$ in an accretion disk having $\dot{m}_{out}=10\dot{m}_{crit}$ around a stellar BH of mass $M_{BH}=10M_{\odot}$.
Increasing the $\alpha$ parameter by a factor of 10, decreases the total luminosity and the real mass accretion rate by approximately $10\%$. 

 The second uncertainty is the exact structure of the porous atmosphere which determines the value of the effective opacity. The effective opacity is given by eq. \ref{eq:kappa_eff}, where $\mr{B}$ is a free parameter, and $\mr{A}$ is determined through eq. \ref{eq:A}. For the same disk as before, we examined the influence of different values of $B$ on the disk structure and luminosity.
The values of $\mr{B}$ taken are $\mr{B}=0.5$, $\mr{B}=1.0$ and $\mr{B}=1.5$. In this case, increasing $\mr{B}$ by a factor of 2 (i.e., decreasing $\kappa_\mr{eff}$), causes a small increase of $6\%$ in the real accretion rate and a modest increase of $15\%$ in the total luminosity. 

\begin{figure}
\centerline{\epsfig{file=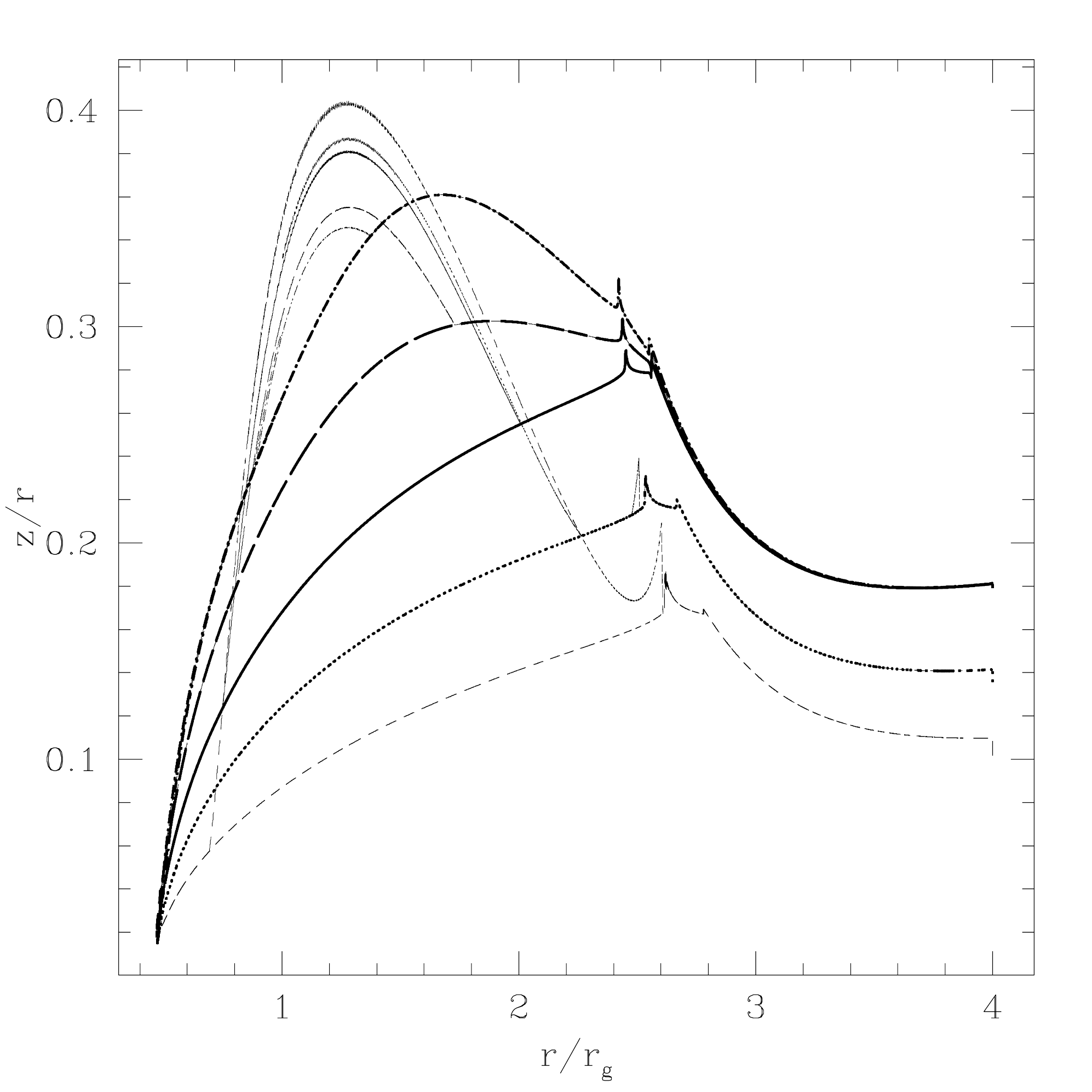,width=8cm,angle=0}}
\caption{Comparison between the vertical structure, $z/r$, of disks with different atmospheric effective opacities and different $\alpha$, for $10\dot{m}_{crit}$ around $M_{BH}=10M_{\odot}$. Solid, dotted and short dashed lines denote  $\alpha=$ 0.001, 0.01 and 0.1 respectively, while  $B=0.5$. Long dashed and dot short-dashed are for $\alpha=0.001$ and $B=1.0, 1.5$ respectively. Thick lines depict the height of the photosphere when an optically thick wind is absent (at outer radii), and the sonic point (inner radii). Thin lines mark the height of the photosphere when present in a thick wind.}
\label{fig:zparameter}
\end{figure}

\begin{figure}
\centerline{\epsfig{file=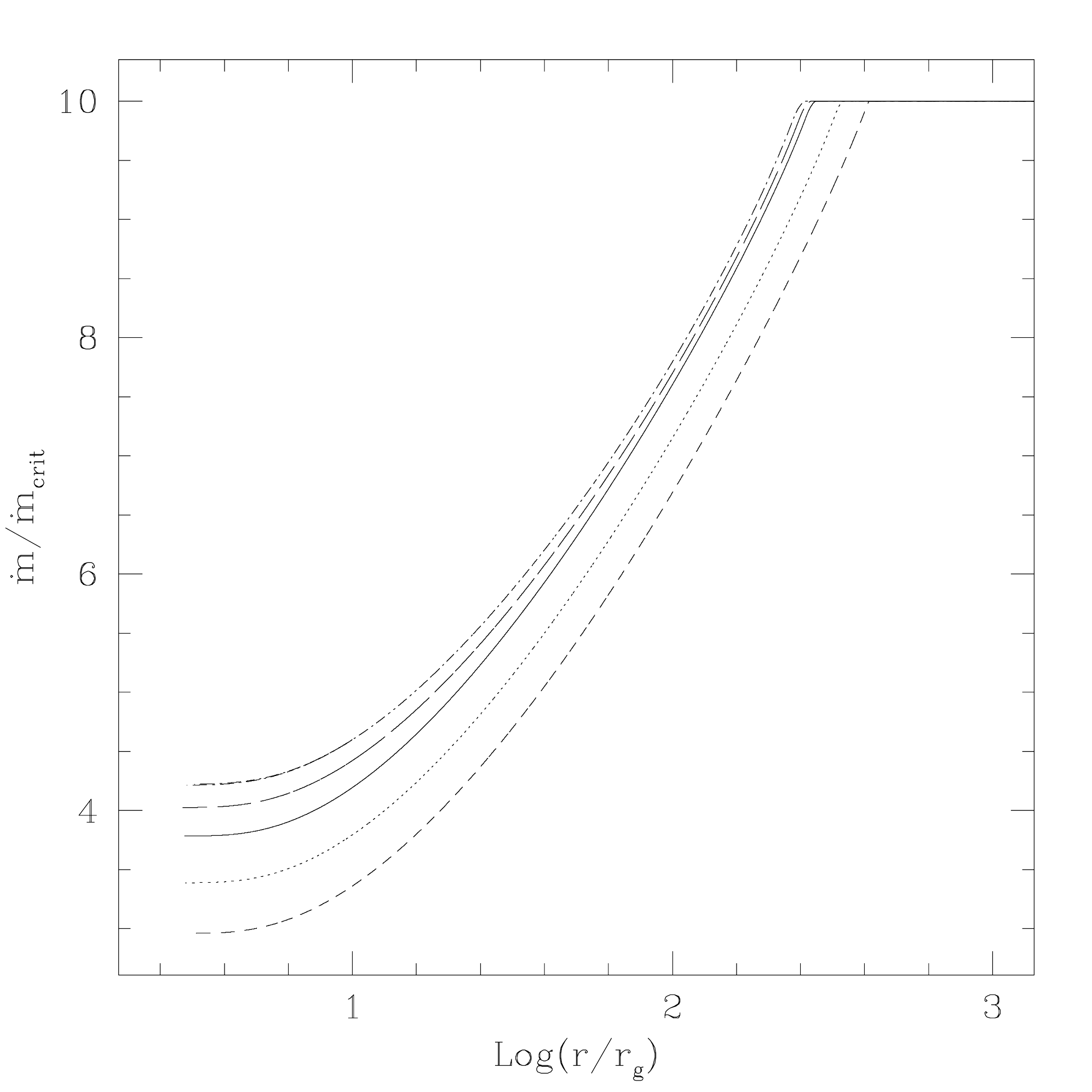,width=8cm,angle=0}}
\caption{Comparison between the accretion rates of disks with different atmospheric effective opacity parameters, $B$, and different viscosity parameters $\alpha$, as labeled in fig.\ \ref{fig:zparameter}. Note that increasing $\alpha$ decreases $\dot{m}_{real}$ while increasing $B$ increases $\dot{m}_{real}$, but the overall effect is not very large.}
\label{fig:mparameter}
\end{figure}

\begin{figure}
\centerline{\epsfig{file=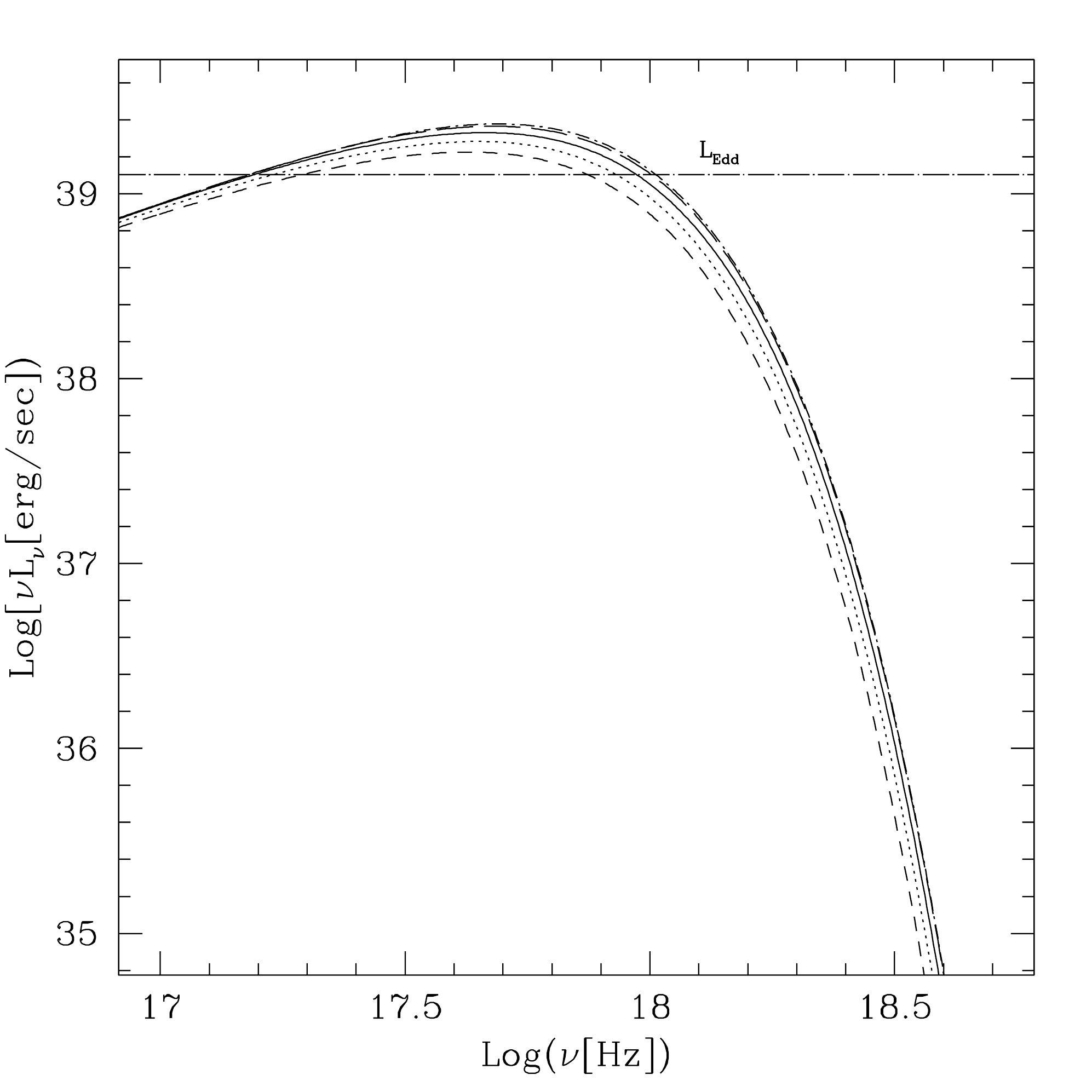,width=8cm,angle=0}}
\caption{Comparison between the spectra, $\log({\nu}L_{\nu}[erg/sec])$ vs. $\log(\nu[Hz])$, for the different conditions given in fig.\ \ref{fig:zparameter}. }
\label{fig:Lparameter}
\end{figure}

In summary, the theoretical uncertainties in the model do not translate into large uncertainties in the model predictions. Figs. \ref{fig:zparameter}-\ref{fig:Lparameter} summarize the aforementioned results in more detail.  

\subsubsection{Super Massive BH}
In addition to accretion disks around a stellar BH, we also examined super critical accretion around a super massive BH of mass $10^{6.5}M_{\odot}$, with outer accretion rates of $10\dot{m}_{crit}$ and $20\dot{m}_{crit}$. 

We find that the real accretion rates for these disks are slightly smaller when compared to the stellar BH case, $2.6\dot{m}_{crit}$ and $3.8\dot{m}_{crit}$ respectively. The relative vertical height ($z/r$) is smaller as well (reaching a maximum value of $z/r\approx0.55$ for the $\dot{m}_\mr{out}=20\dot{m}_{crit}$, as compared to  $z/r\approx0.65$ in the equivalent stellar BH case. 

\begin{figure}
\centerline{\epsfig{file=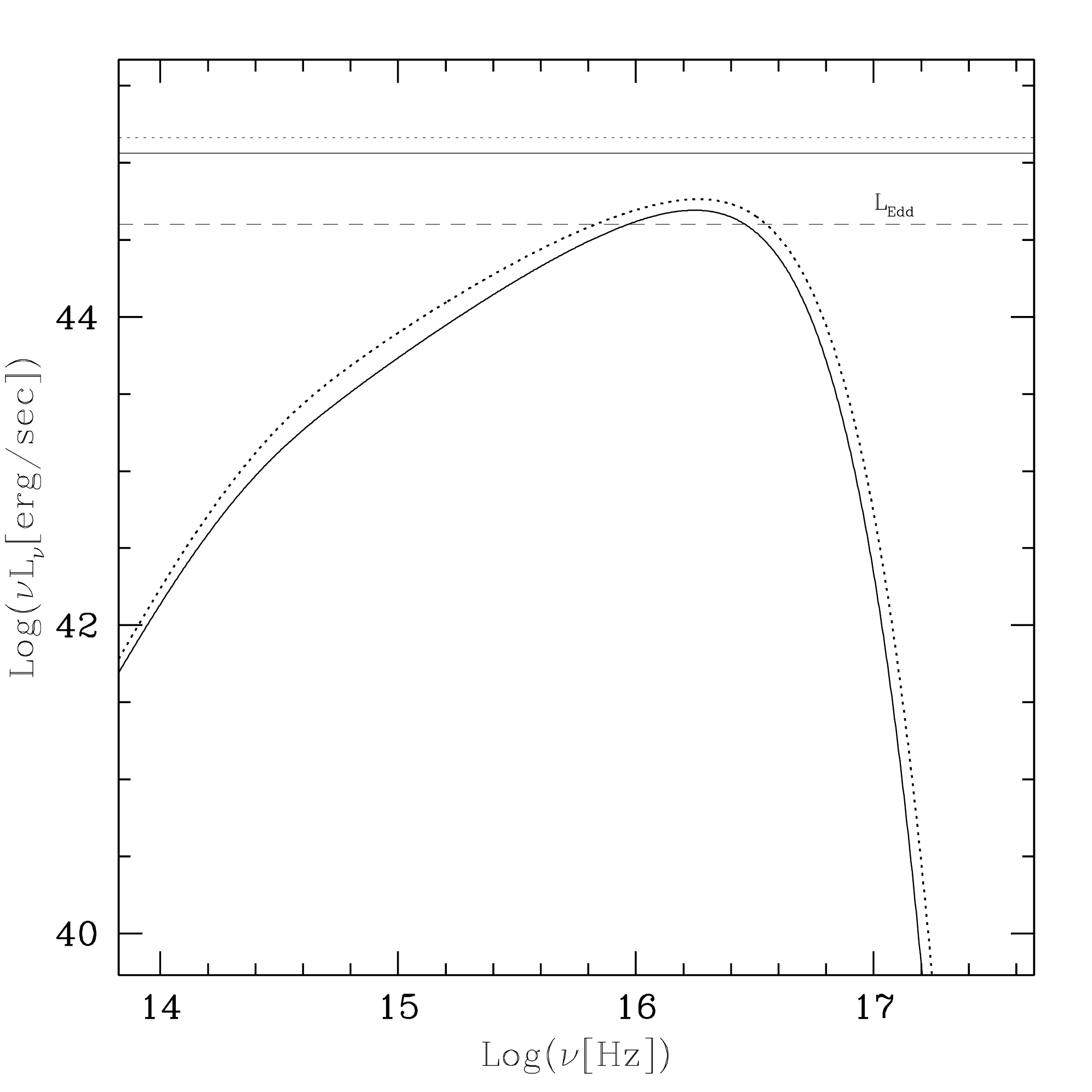,width=8cm,angle=0}}
\caption{The emitted spectra, $\log({\nu}L_{\nu}[erg/sec])$ vs.\ $\log(\nu[Hz])$, by accretion disks onto a super-massive BH, with a mass of $10^{6.5}M_{\odot}$. The outer accretion rates are $10\dot{m}_{crit}$ (solid line) and $20\dot{m}_{crit}$ (dotted line). The horizontal thin lines are the total luminosities and dashed line is Eddington luminosity for comparison. }
\label{fig:lsupermassive}
\end{figure}

The main difference is in the emergent spectra. The spectra of disks around massive BHs has a maximum at lower frequencies (far-UV compared with the x-ray for stellar BHs, see fig.\ \ref{fig:lsupermassive} for comparison). The total luminosities are $2.9L_{Edd}$ and $3.7L_{Edd}$ respectively, i.e., modestly less super-Eddington.

\begin{table*}
\begin{tabular}{c c c c c}
\hline
$\dot{\mr{m}}_{\mr{out}}/\dot{\mr{m}}_{\mr{crit}}$ & $\dot{\mr{m}}_{\mr{real}}/\dot{\mr{m}}_{\mr{crit}}$ & $\mr{L}_{\mr{tot}}/\ledd$ & Energy Dissipation & efficiency ($\eta/\eta_0$)\\
\hline
1 & 0.74 & 0.79 & 0.8 & 1.05\\
5 & 2.4 & 2.65 & 2.75 & 1.1\\
10 & 3.8 & 3.9 & 4.45 & 1.03\\
20 & 5.7 & 4.85 & 6.9 & 0.85\\
\hline
\end{tabular}
\caption{Summary of the slim accretion disks around $10M_{\odot}$ with $\alpha=0.001$. The mass accretion rates are given in units of the critical rate while the luminosity and energy dissipation (which is the total energy produced by viscosity) are given in units of the Eddington luminosity of the BH. The last column is the efficiency of the disk compared to the standard disk efficiency (see eq.\ \ref{eq:mdotcrit}), calculated with the {\em real} accretion rate. The efficiency calculated using the {\em outer} accretion rate is smaller. At high accretion rates, a significant wind is accelerated, taking some of the energy and reducing the efficiency.}
\label{tab:results}
\end{table*}

\section{Discussion}
\label{sec:discussion}


In the present analysis, we  searched numerically for possible solutions describing super-critical accretion disks around black holes, while allowing for the formation of ``porous" layers with a reduced opacity. We found solutions with significantly super-critical accretion rates, in which the vertical disk height is  smaller than the radius, that is, {\em slim} disks. 
Because the super-Eddington state excites a strong wind, the actual mass accretion onto the BH can be notably smaller. Solutions were found to exist with accretion rates ranging between about 0.5 ${\dot m}_{crit}$ to about 20${\dot m}_{crit}$. In all cases, there is a photon tired continuum driven wind.

At the low range, the disks are overall sub-Eddington, but locally the flux can surpass the critical value, and therefore it can excite continuum driven winds.  Namely, the critical accretion rate given by eq.\ \ref{eq:mdotcrit} is not the lower limit for wind generation. At the low range, almost all the energy dissipated is either radiated away or transferred to the wind. Moreover, the wind is then optically thin and the photosphere coincide with the sonic point of the wind.

As the outer mass accretion rate increases, the winds become more massive, thereby reducing the fraction of mass accreted onto the black hole. Also, some of the dissipated energy is then advected with the flow into the BH.   

For very high accretion rates surpassing about 20$\dot{m}_{crit}$, the photosphere which resides in the wind is found to be located at $z \gtrsim r$. This implies that the 1D+1D type of solution described here breaks down. Instead, one has to look for a solution in which the semi-hydrostatic inflow has a disk-like solution, while the super-sonic outgoing wind has a spherical-like solution.  The description of such extremely high accretion rate disks is the subject of a future publication.  

One of the uncertainties in the model is the opacity law behaviour of the porous atmosphere. The reasons it is not known well is because it depends on the nonlinear radiative hydrodynamic configuration the atmosphere will reach, and unfortunately, there are still no numerical simulations or empirical data which can constrain the effective opacity law. It is for this reason that we parameterized the effective opacity (see eq.\ \ref{eq:kappa_eff})

 As can be seen in figs.\ \ref{fig:mparameter}-\ref{fig:Lparameter} almost all disks characteristics have either a small or a modest sensitivity to the changes in the opacity law. For example $\dot{\mr{m}}_{\mr{real}}/\dot{\mr{m}}_{\mr{out}}$ varies between 0.43 to 0.44, or the total luminosity varies between $\sim 0.570 \ledd$ to $0.576 \ledd$, while changing the opacity parameter $B$ by 50\%. This implies that the uncertainties do not undermine the model predictions. On the other hand, it would be impossible to use super-Eddington accretion disks to constrain the relevant parameters. 
 
 \section{Acknowledgements}
N.J.S. is grateful to the support of ISF grant 1325/06. 

\bibliography{thebibfile}{}
\bibliographystyle{mn2e}

\bsp
\label{lastpage}
\end{document}